\documentclass[nofootinbib,twocolumn,floats,aps]{revtex4}
\usepackage{epsfig}

\usepackage{bm}
\usepackage{longtable}
\def\Rb {{\bf R}}
\def\Tb{ {\bf T}}

\def\bb {{\bf b}}

\def\sixj#1#2#3#4#5#6{\left\{\negthinspace\begin{array}{ccc}
#1&#2&#3\\#4&#5&#6\end{array}\right\}}
\def\M {{{\cal M}}}
\def\sqi{\frac{1}{\sqrt{2}}}
\def\x{\times}
\def\nn{\nonumber }

\def\endauthors{}
\def\authors#1\endauthors{#1}
\def\fot{\frac{1}{2}}

\def\mbt{\mbox{\boldmath$\tau$}}

\def\rf#1{{(\ref{#1})}}

\def\bra#1{\langle #1|}
\def\ket#1{|#1 \rangle}
\def\Ket#1{||#1 \rangle}
\def\Bra#1{\langle #1||}
\def\rb {{\bf r}}
\def\pb {{\bf p}}
\def\Pb{ {\bf P}}
\def\be{\begin{equation}}
\def\ee{\end{equation}}
\def\br{\begin{eqnarray}}
\def\er{\end{eqnarray}}

\def\binn#1#2{\left\{\negthinspace\begin{array}{l}#1\\#2\end{array}\right.}

\def\jN{{\sf{j}}_N}
\def\jL{{\sf{j}}_\Lambda}

\begin{document}
\title{Evaluation of coincidence number of pairs and asymmetry parameter in nonmesonic hypernuclear decay}

\author{C. Barbero$^{1,2}$}
\author{A. Mariano$^{1,2}$}
\author{S. B. Duarte$^{3}$}

\affiliation{$^1$Facultad de Ciencias Exactas
, Departamento de F\'isica, Universidad Nacional de La Plata, 1900 La Plata,
Argentina}

\affiliation{$^2$Instituto de F\'isica La Plata, CONICET, 1900 La
Plata, Argentina}

\affiliation{$^3$
Centro Brasileiro de Pesquisas F\'{\i}sicas, Rua Dr Xavier Sigaud 150\\
CEP 22290-180, Rio de Janeiro-RJ, Brazil}

\date{\today}

\begin{abstract}

We develop  a theoretical framework for the evaluation of
the coincidence number of pairs, $N_{nN}$, and the asymmetry parameter, $a_\Lambda$,
in nonmesonic hypernuclear decay $\Lambda N\rightarrow nN$.
The primary nomesonic weak hyperon-nucleon decay process is described through the
independent particle shell model. When the effect of the strong
interactions experimented by the two nucleons leaving the residual
nucleus is included, we have in addition to the
$2p1h$ primary configuration $2p'1h'$, $3p2h$, $4p3h$, $\cdots$, etc secondary ones.
We work within the $2p'1h'$ subspace of final states assuming a simple
approach for finite nuclei based on the eikonal approximation.
An optical potential including the
nucleon-nucleus isoscalar and isovector interaction is introduced to take into account
the minimum nuclear medium effects and to describe the
nucleon-nucleus dispersion process along the outgoing path.
We applied the proposed description to the calculation of the
observables in $^{5}_\Lambda He$ and $^{12}_\Lambda C$ nonmesonic
decays. The calculated results show that our treatment of FSI even
considered as restricted to a subspace of states gives
a good agreement with experimental data, similar to that ones presented by more
elaborated evaluations. This indicates that
the developed theoretical scheme seems to be appropriated to include
more complex configurations.  We also discuss the ground state normalization effects at the moment of including the two-nucleon induced decay ($\Lambda NN \rightarrow nNN$)
contribution.

\bigskip

{\hspace{-.35cm}PACS numbers: 21.80.+a, 25.80.Pw, 21.60-n, 13.75.Ev}

\end{abstract}

\maketitle

\section{Introduction}
\label{intro}

Although a free $\Lambda$ particle decays through the mesonic mode $\Lambda \rightarrow \pi N$, inside nuclear medium this channel is Pauli blocked and the nonmesonic hypernuclear weak decay (NMHD) $\Lambda N\rightarrow n N$ emerges as the dominant channel. This primary decay can be induced by a neutron ($\Lambda n\rightarrow nn$) or a proton ($\Lambda p\rightarrow np$), with widths $\Gamma_n$ and $\Gamma_p$, respectively. During the last three decades a high effort to solve the commonly called nonmesonic hypernuclear decay puzzle has been done. The controversy is associated to the theoretical troubles in reproducing simultaneously the measured values of: i) the detected number of neutrons,
protons, $nn$ and $np$-pairs, $N_{n}$, $N_{p}$, $N_{nn}$ and $N_{np}$, respectively  (or, equivalently, the spectra of emitted protons as well as $nn$ and $np$-coincidence spectra); ii)
the intrinsic asymmetry parameter, $a_\Lambda$, which is determined by the interference terms between the parity-conserving (PC) and parity-violating (PV) transitions to final states with
different isospins. These observables lead information about both, the primary nonmesonic decay and the subsequent final state interactions (FSI) originated by the nuclear medium.

Since the pioneering works of Block and Dalitz \cite{Blo63} and Adams \cite{Ada67} many theoretical attempts are trying to improve the description of NMHD, such as:

i) modifications of the exchange potential incorporating mesons heavier than pion;

ii) inclusion of correlated and uncorrelated two-pion exchange;

iii) inclusion of interaction terms that violate the $\Delta T=\fot$ isospin rule;

iv) addition of quark degrees of freedom;

v) analysis of the contribution of two nucleon (2N) induced decay mode $\Lambda NN\rightarrow NNN$;

vi) evaluation of the effect of FSI between the outgoing primary nucleons and those in the residual nucleus, etc.

Particularly, in Refs. \cite{Ra97,Gar03,Alb04,Chu0708} the authors show that FSI are an important ingredient when studying this decay mode. Indeed, it was shown that the value of the neutron to proton ratio, $\Gamma_n/\Gamma_p$,
extracted from the experiments and also the intrinsic asymmetry parameter are strongly modified by the FSI effects. Actually, there is a general consensus in respect to the essential role of FSI to reach a realistic calculation of NMHD observables. Among several models used for evaluating the effect of FSI we mention: i) those based in the intranuclear cascade code (INC) \cite{Ra97,Gar03,Pa01,Gar04,Gar05}, and ii) a microscopical model developed within a nuclear matter formalism \cite{Bau07}; iii) models describing the interaction between the outgoing primary particles and the residual core by means of an optical potential \cite{Ram92,Con09}. In the first case the nucleon propagation inside the residual nucleus is simulated by a Monte Carlo code, in which the nucleons produced in the weak decay are followed through a semi-classical path until they leave the nucleus. In the second one, the FSI are included within a nuclear matter picture with the help of the local density approximation in an effective way via the first order term of the residual interaction in the RPA amplitude. Finally, evaluations using optical potentials describe the average interaction felt by the outgoing particles within an effective scheme.

Motivated by the previous discussion, we want to develop a finite nucleus theoretical frame that enables an evaluation of $N_{nN}$, $N_{nn}/N_{np}$ and,
simultaneously, $a_\Lambda$, which should be an extension of the of our previous shell model calculations \cite{Bar08}. By simplicity we begin making an evaluation of the mixing between $2p1h$ final configurations,
in the simpler way working within the eikonal approximation.
It provides a clear description of the minimum nuclear
medium action on the emerging nucleons through an isospin dependent optical potential, not yet considered in previous
works \cite{Ram92,Con09}. The paper is organized as follows: the formalism for the evaluation of the
observables and the inclusion of the nucleon-nucleus interaction via the eikonal approximation is
presented in Sect. 2; numerical results are exhibited in Sect. 3 and concluding remarks are briefly drawn in Sect. 4.

\section{Formalism}

\subsection{The primary decay model}

The partial decay rate, $\Gamma_N$, for the primary $\Lambda N\rightarrow n N$ nonmesonic decay of an initial hypernucleus (with spin $J_IM_I$ and energy ${\cal E}_I$) to a residual nucleus (with spin $J_FM_F$ and energy ${\cal E}_F$) plus two free nucleons (with spins, isospins and momenta $s_i$, $t_i$ and $\pb_i$, respectively, being $i=n,p$) and the intrinsic asymmetry parameter, $a_\Lambda$, are evaluated from the expressions \cite{Bar02,Bar03,Bar05}
\begin{eqnarray}
\Gamma_{N}&=&\int d\Omega_{p_n}\int d\Omega_{p_N}\int dF_{N}\sum_{s_ns_Nt_nt_NM_F}\nn\\
&\x&|(\pb_ns_nt_n\pb_Ns_Nt_N;J_FM_F|V|J_IM_I)|^2,
\label{a}\\
a_\Lambda& =& \frac{3}{\sum_{M_I}\sigma(J_IM_I)}\nn\\
&\x&
\binn{\hspace{0.cm}\frac{1}{J_I+1}\sum_{M_I}M_I\sigma(J_IM_I)\hspace{0.9cm}
\mbox{for $J_I=J_C+\frac{1}{2}$},}
{-\frac{1}{J_I}\sum_{M_I}M_I\sigma(J_IM_I)\hspace{1cm}\mbox{for $J_I=J_C-\frac{1}{2}$},}
\nn\end{eqnarray}
where
\br
\sigma({J_IM_I})&=&\int d\Omega_{p_p}\int dF_p\sum_{s_ns_pM_F}\nn\\
&\x&|(\pb_ns_nt_n\pb_ps_pt_p;J_FM_F|V|J_IM_I)|^2,
\label{d}\er
with
\br
\int dF_{N}&=&2\pi\sum_{J_F}\int\frac{p_N^2dp_N}{(2\pi)^3}\int\frac{p_n^2dp_n}{(2\pi)^3}\label{e}\\
&\x&\delta\left(\frac{p_n^2}{2m_N}+\frac{p_N^2}{2m_N}+\frac{|\pb_n+\pb_N|^2}{2m_N(A-1)}-\Delta_{F}
\right).\nn
\er
Here $\Delta_{F}={\cal E}_I-{\cal E}_F-2m_N$ is the released energy with $m_N$ being the nucleon mass,
$A$ is the residual nucleus mass number and $V$ is the transition potential used to describe the primary nonmesonic decay process. We have employed the one-meson-exchange model from Refs. \cite{Bar02,Bar03,Bar05} which
uses a standard strangeness-changing weak $\Lambda N\rightarrow
NN$ transition potential comprising the exchange of the complete
pseudoscalar and vector meson octets ($\pi$, $\eta$, $K$, $\rho$,
$\omega$, $K^*$).

The number of emitted $nN$-pairs after this weak decay is evaluated weighting the number of pairs appearing in a given state $nN'\equiv N'$ ($N^{N'}_{nN}$)
by the probability of reaching that state:
\be
N_{nN}^{wd}=\sum_{N'} N^{N'}_{nN}\frac{\Gamma_{N'}}{\Gamma_{n}+\Gamma_{p}}.
\label{e'}\ee
Note that $N^{n}_{nn}=1$, $N^{p}_{nn}=0$, $N^{p}_{np}=1$ and $N^{p}_{nn}=0$.

Following Refs. \cite{Bar02,Bar03,Bar05}, after changing to relative and center of mass variables
\br
\pb&=&\frac{1}{2}(\pb_N-\pb_n),\hspace{1.5cm}\Pb=\pb_n+\pb_N,\nn\\
\rb&=&\rb_N-\rb_n,\hspace{2.2cm}\Rb=\fot(\rb_n+\rb_N),
\label{8}\er
performing the partial wave expansion of the emitted nucleon waves,
and making the angular momentum recoupling, the observables can be written in terms of the
matrix elements
\br
\bra{plPL\lambda SJT\nu_FJ_F;J_I}V\ket{J_I}&=&
\hat{J}_I^{-1} \sum_{\jN} f_J(\jN,\nu_FJ_F)\nn\\
&\x&\M(plPL\lambda SJT;{\jL
\jN}),\nn\\
\label{45}\er
with $\jL\equiv
n_\Lambda\,l_\Lambda\,j_\Lambda\,t_\Lambda $ and $\jN\equiv
n_N\,l_N\,j_N\,t_N$ being the single-particle states for the lambda and
nucleon, respectively (we assume that the $\Lambda$ particle
behaves as a $\ket{\frac{1}{2},-\frac{1}{2}}$ isospin particle in
the $1s_{1/2}$ level) and $\nu_F$ specifying
the remaining quantum numbers in the final state besides the
nuclear spin. The factors are defined as
\br
f_J(\jN,\nu_FJ_F)&=&(-)^{2J_F}\hat{J}\hat{J}_I\sixj{J_C}{J_I}{j_\Lambda}{J}{j_N}{J_F}\nn\\
&\x&\Bra{J_C}a_{\jN}^\dag \Ket{\nu_F J_F}, \label{5a}\er
with $J_C$ being the core spin and the matrix
element
\begin{eqnarray}
\M(plPL\lambda SJT;\jL \jN) &=&\sqi\left[1-(-)^{l+S+T}\right]\nn\\
&\x&
({plPL\lambda SJT}|V|{\jL \jN} {J}). \label{6a}\end{eqnarray}
Within the independent particle shell model
the core and final nuclear states are described as $|J_C>\rightarrow
|0>$ and $|J_F>\rightarrow |j_b^{-1};J_F>$ for $^5_\Lambda$He, with $|0>$ being the vacum state,
and $|J_C>\rightarrow |j_a^{-1};J_C>$
and $|J_F>\rightarrow |j_a^{-1}j_b^{-1};J_F>$ for $^{12}_\Lambda$C.

The effect of final state
interactions between the two outgoing $nN$ particles, which has been
extensively discussed in the literature and is
known to be very important \cite{Pa01}, is
included phenomenologically by modifying the two emitted nucleons plane waves as in our mentioned works: short range
correlations are treated at a simple Jastrow-like level multiplying the exchange potential by the correlation function
\be
g_{NN}(r)=1-j_0(q_c r),
\label{222}\ee
with $q_c=3.93$ fm$^{-1}$, while the
contribution of finite nucleon size effects at the interaction
vertices are gauged by a monopole form factor
$(\Lambda_M^2-\mu_M^2)/(\Lambda_M^2+q^2)$,
being $\Lambda_M$ the cutoff for the meson $M$ \cite{Bar02}. Additionally,
corrections due to kinematical effects
related to the $\Lambda$-nucleon mass difference and the
first-order nonlocal terms are taken into account \cite{Bar03}.

\subsection{Final state interactions and ground state correlations}

When one of the ejected nucleons interacts with the remaining
ones in the residual nucleus, the $2p1h$ primary configuration
$|N) \equiv |\pb_ns_nt_n\pb_Ns_Nt_N;J_FM_F)$
(where a hole $N^{-1}$ produced on the initial nucleus by promotion
of the particle $N$ is present in the final state) can be mixed with
other $|N')$ configurations and also
with  more complex ones as $3p2h$, $4p3h$, $\cdots$.
This is schematically shown in Fig. 1.
By simplicity, let's consider $\{|2p1h),|3p2h)\}$ as the final space
(note that $3p2h$ configurations participate in the 2N mode).
The $3p2h$ state will be
$|NN')\equiv |\pb_ns_nt_n\pb_Ns_Nt_N  \pb_{N'}s_{N'}t_{N'};J_FM_F)$ ,
with $N$ and $N'$ being $n$ or $p$ (note that an additional hole $N'^{-1}$
is produced in the final nucleus).
Now, Eqs. \rf{a}-\rf{e} could be straightforward modified to evaluate the decay
rate to a state $|NN')$, $\Gamma_{N N'}$, enlarging the final phase space, adding the corresponding integrals, and changing $A-1$ by $A-2$.
The number of emitted $nN$-pairs
in Eq. \rf{e'} should then be expressed as
\be
N_{nN}= \sum_{N'} N^{N'}_{nN}
\frac{\Gamma_{N'}}{\Gamma_T}
+ \sum_{N' N''} N^{N' N''}_{nN}
\frac{\Gamma_{N'N''}}{\Gamma_T}
,\label{e''}\ee
where $N^{N' N''}_{nN}$ represents the number of $nN$ pairs appearing in a given $nN' N''\equiv N' N''$ state
and $\Gamma_T\equiv\Gamma_{n}+\Gamma_{p}+\Gamma_{nn}+\Gamma_{np}+\Gamma_{pp}$
is the total rate. Note that $N^{nn}_{nn}=3, N^{np}_{nn}=1, N^{pp}_{nn}=0, N^{pp}_{np}=2, N^{np}_{np}=2, N^{nn}_{np}=0$. Because $V$ is a two-body operator and as for the moment we are not considering ground state correlations (GSC) we have $(N N'|V|J_IM_I)=0$, giving $\Gamma_{N N'}=0$. This is because the mixing
originated by FSI is not included in the unperturbed $\{|N),|NN')\}$ basis used in Eq. \rf{e''}. Instead of this,
we should use  a perturbed basis $\{\widetilde{|N)},\widetilde{|NN')}\}$ in the valuation of $\Gamma_{N}$ and
$\Gamma_{N N'}$, with
\br
\widetilde{|N)} &=& c_N |N) + \sum_{N'} c_{N,N'} |N') + \sum_{N'N''} c_{N,N'N''}  |N'N''),\nonumber\\
\widetilde{|NN')} &=& c_{NN'} |NN') + \sum_{N''N'''} c_{NN',N''N'''} |N''N''')\nonumber\\
 &+& \sum_{N''} c_{NN',N''}  |N''),
\label{ef}\er
where the coefficients $c_{i}$ and $c_{i,j}$ should be evaluated in the stationary perturbation theory, at least at
the second order to take properly into account normalization effects \cite{Mariano94b}.
The fact of using a perturbative scheme favors the identification of the numbers of pairs coming from each final state.
Now, the $|NN')$ configurations are reached by the primary decay to the states $|N'')$ present
in the last term of Eq. \rf{ef}, and by the coupling of these states with the $|NN')$ one produced
by the strong interaction responsible of the FSI which enters in the evaluation of the coefficient $c_{NN',N''}$.
The number of pairs can be calculated now from Eq. \rf{e''} replacing
$\Gamma$ by $\widetilde{\Gamma}$, being $\widetilde{\Gamma}$ calculated through Eq. \rf{a} by using the perturbed basis.
Because at this stage it is not important the exact energy position of the states, we can take the unperturbed ones for the evaluation.

\begin{figure}[h!]
\vspace{-3.5cm}
\begin{center}
\includegraphics[height=12.cm,width=8.5cm]{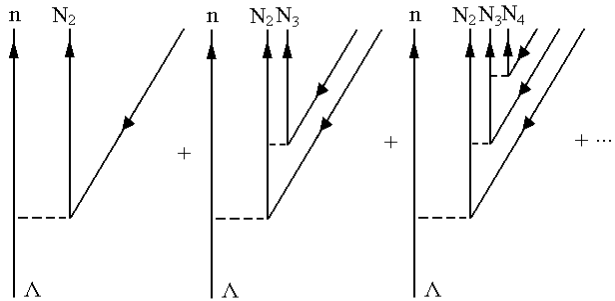}
\end{center}
\vspace{-4.cm}
\label{fig:0} \caption{The left diagram schematizes the primary decay that leads to a $2p1h$ final state and the additional ones represent the action of the FSI to produce more complicated configurations.}
\end{figure}

Finally, we mention that if $2p2h$  GSC were admitted in the ground state \cite{Mariano94}, we could write
\br
\widetilde{|J_IM_I)} &=& c_0 [|J_IM_I) + \sum_{2} c_{2} |2 J_IM_I)],
\label{efg}\er
where with $"2"$ we indicated a $2p2h$ configuration. Now the coefficient $c_0$ reduce the $\widetilde{\Gamma}$ contributions mentioned
previously, and the $2p2h$ contributions give place to the 2N mode. One must take care the way $c_0$ and $c_2$ are evaluated \cite{Mariano96}.
If we use  a first order perturbation theory we must take,
\be
c_0= 1,~~~c_2 = {(2J_IM_I|V_s|J_IM_I) \over  - E_2}, \label{efgh'}
\ee
where $V_s$ indicates the nucleon-nucleon strong residual interaction. We remark that it is not correct to take
\br
c_0 = \sqrt{{1 \over 1 + \sum_2|c_2|^2}} \simeq 1 - 1/2 \sum_2|c_2|^2 + \cdot \cdot \cdot,\label{efgh}\er
because, except for the first term, the others lead to an infinite series of disconnected graphs in the amplitude, and it is well known this is unphysical and these contributions can be canceled enlarging the space by the inclusion of $4p4h$ in the ground state \cite{Mariano96}.
In addition, the norm correction included as in Eq. \rf{efgh} leads to a nonextensivity of nuclear matter introducing an $A$ dependence, which is
a very strong drawback of the approximation \cite{Mariano96}.  It would be more appropriate to make
a direct diagonalization in the $\{|J_IM_I),|2 J_IM_I)\}$ basis and take
$(2 J_IM_I|V_s|2' J_IM_I)\simeq 0$ because of the high density of these configurations. This
leads to
\be
c_2 = {(2J_IM_I|V_s|J_IM_I) \over E_I - E_2},
\ee
where $E_I$ is now solution of the secular equation \cite{Mariano94}
\be
E_I = \sum_2{|c_2|^2 \over E_I -E_2},
\ee
and $c_0$ is obtained by normalization. This was done in a finite nucleus frame for the case of $^{48}Ca$ using
the MY3 force \cite{Mariano94} getting a value of $c_0= 0.69$, which could be enlarged for the lighter
nucleus we interested in here due to the smaller $2p2h$ configurations space.

\subsection{The minimum effect of nucleon-nucleus interaction}

In order to see how FSI effects change the shell model results we analyze the effect of mixing between different
$|N)$ configurations in the final states. For this purpose we keep only the first formula and the first two terms of it in Eq. \rf{ef}.
In the present work instead of making an explicit evaluation of the coefficients $c_N$ and $c_{N,N'}$ in a second order perturbation theory,
we are proposing a treatment of FSI by using the  simplified final nucleon states constructed in the framework of the nonrelativistic eikonal approximation \cite{McCau58,Joa75,Ama83,Deb00,Landau}. We remember that in the energy regime of a few hundred of MeV the nucleon-nucleus scattering is satisfactorily described by this approach (see, for example, applications to A(p,pN) reaction in Refs. \cite{Ove06,Jac66,Kull71,Ben78}). We have to bear in mind that the available energy in primary process leaves a momentum $p_N\sim 400$ $\mbox{MeV}/{c}$ for each one of these nucleons. To construct the nucleon final states of interest we start from free nucleons state inside the nucleus and study their modification in crossing the nuclear medium (represented by an optical potential) within the eikonal approximation. The resulting state of this nuclear medium scattering process is taken as the final nucleon state for the NMHD calculation. Here is very important to mention that, within this simple model, we can not include the contribution of secondary nucleons produced by the strong interaction, which largely affect the numbers $N_{nn}$ and $N_{np}$ (second formula in Eq. \rf{ef}). This is the reason by which we have remarked that we are including the minimum effect of nucleon-nucleus interaction corresponding to the scattering of the outgoing particles. Nevertheless, in this work we will analyze the role played by the isovector nucleon-nucleus interaction in the scattering of the emerging particles, not included in previous optical model treatments.
Additionally, because the dependence of the ratio $N_{nn}/N_{np}$ with the 2N contribution is moderate \cite{Gar03,Gar04} we not consider its effect here.

In coordinate representation the accumulated effect of all the interactions of the emerging nucleons along their outgoing path inside the residual nucleus is incorporated through the action of an operator factor $S^\dagger(\rb)$ on the unperturbed plane wave functions as shown in Ref. \cite{Ove06}. Concretely, the outgoing nucleons final state is obtained as
\br
&&<\rb_n | \pb_n> <\rb_N | \pb_N>\nn\\
&=&S^\dagger(\rb_n,\pb_n)S^\dagger(\rb_N,\pb_N) e^{i\pb_n\cdot\rb_n}e^{i\pb_N\cdot\rb_N},
\label{1}\er
where $\rb_N$ is the coordinate of the nucleon and
\be
S(\rb,\pb)=
e^{-i\frac{m_N}{p}\int_{z_{p}}^{+\infty}dzV_{\mbox{\tiny{opt}}}(\bb,z)}.
\label{2}\ee
We have chosen the momentum $\pb$ along the $z$ axis, with $\rb=(\bb,z_p=\hat{\pb}\cdot\rb)$ representing the nucleon location after the primary nomesonic decay and $(\bb,z)$ the posterior collision points, respectively. The nucleon is crossing the nucleus with an impact parameter $\bb$, with $V_{\mbox{\tiny{opt}}}(\rb)$ being the optical potential associated to the nuclear medium representation. We adopt the simplified form \cite{Boh68}:
\be
V_{\mbox{\tiny{opt}}}(\rb)=\binn{-\left(V_{0}+\frac{V_1}{A}
\mbt\cdot\Tb\right)\hspace{0.5cm}\mbox{if}\hspace{0.5cm}|\rb|\leq R}
{0\hspace{3cm}\mbox{if}\hspace{.5cm}|\rb|> R}, \label{3}\ee
with $R=1.25A^{1/3}$ fm  being the nuclear radius, $V_0$ and $V_1$ representing the real isoscalar and isovector "depths", respectively, of the optical potential. The latter part depends on the nucleon and residual nucleus isospin operators, $\mbt$ and $\Tb$, respectively. Integration in the $z$-variable allows to write the factor $S(\rb_N,\pb_N)$ as
\br
S(\rb_N,\pb_N)&=& e^{i\frac{m_N}{p_N}\left(V_{0N}+\frac{V_{1N}}{A}\mbt_N\cdot\Tb\right)
\sqrt{R^2-b^2}}\nn\\
&\x&e^{-i\frac{m_N}{p_N^2}
\left( V_{0N}+\frac{V_{1N}}{A}\mbt_N\cdot\Tb\right)\pb_N\cdot\rb_N},
\label{4}\er
with $V_{0N}$ and $V_{1N}$ being the isoscalar and isovector nucleon potential depth. Note that, in order to simplify the changes to the two-nucleon center of mass variables, and to perform integrals analytically in Eqs. (1-3), we can neglect the energy dependence of the optical potential parameters (a rather good approximation in the energy range of interest) and take the same values of isoscalar and isovector depths for protons and neutrons:
$V_{0n}\simeq V_{0p}\simeq \tilde{V}_{0}$ and $V_{1n}\simeq V_{1p}\simeq \tilde{V}_{1}$. In addition, when using the result of Eq. \rf{4}, the quantities $p_N^2$ and $b^2$ will be substituted by average values. For the first we take
\be
p_N^2\simeq <p>^2=\left(\frac{p_F+p_{\mbox{\tiny{max}}}}{2}\right)^2 ,
\ee
where $p_F=\sqrt{2m_N\epsilon_F}$ and  $p_{\mbox{\tiny {max}}}=\sqrt{2m_N\Delta_F\left(\frac{A-1}{A}\right)}$ are the Fermi and maximum momentum allowed kinematically to nucleons, respectively, being $\epsilon_F$ the Fermi energy. For the second we assume,
\be
b^2\simeq <b^2>\simeq \frac{2}{3}<r^2>=\frac{2}{5}R^2.
\ee
In summary, the effects of FSI are included in the calculation by means of the replacement
\br
e^{i\pb_n\cdot\rb_n}e^{i\pb_N\cdot\rb_N}&\rightarrow&
e^{-2i\frac{m_N\tilde{{V}}_{0}}{<p>}\sqrt{R^2-<b>^2}}\label{5}\\
&\x&e^{i\left(1+\frac{m_N}{<p>^2}\tilde{{{V}}}_{0}\right)\left(\pb_n\cdot\rb_n+\pb_N\cdot\rb_N\right)}\nonumber\\
&\times& e^{-i \frac{m_N}{<p>^2}\frac{\tilde{V}_1}{A}(<p>\sqrt{R^2-<b>^2} - \pb_n\cdot\rb_n)\mbt_n\cdot\Tb}\nonumber\\
&\times& e^{-i \frac{m_N}{<p>^2}\frac{\tilde{V}_1}{A}(<p>\sqrt{R^2-<b>^2} - \pb_N\cdot\rb_N)\mbt_N\cdot\Tb}.
\nn\er
To evaluate the action of the isospin operators, we express them as
\be
\mbt_N\cdot\Tb=(t_+)_NT_-+(t_-)_NT_++2(t_0)_NT_0,
\label{6}\ee
and expand the exponential $e^{i\beta_N\mbt_N\cdot\Tb}$ in power series. Since the operator $\Tb$ acts on the residual nucleus state, this will establish a difference between the calculation of final nucleon state imposed by different isospin state of the decaying hypernucleus, as we will see latter for $^5_\Lambda$He and $^{12}_\Lambda$C hypernuclei. In fact, in the first case the residual core has a half integer  isospin with projections $\pm\fot$, whereas the second hypernucleus has an integer isospin with projections $-1,0$. Additionally, is important to remark that the operators $t_\pm$ allow the possibility of charge exchange. This will produce final neutrons (protons) from the primary proton (neutron) induced decay, which will lead to modifications in the obtained values for the observables, specially the number of detected particles. Leaving all this considerations into account,
the effect of FSI can be incorporated within our formalism by means of the replacement
\br
e^{i\pb_n\cdot\rb_n}e^{i\pb_N\cdot\rb_N}&\rightarrow&
e^{i \alpha_N\pb\cdot\rb}e^{i \alpha_N\Pb\cdot\Rb}\hspace{0.1cm}\mbox{for $N=n,p$,}
\label{7}\er
where we have defined
\br
\alpha_N&=&\left(1+\frac{m_N}{<p>^2}\tilde{{V}}_{N}\right).
\label{9}\er
It is important to specify here the effective strength dependence on the residual nucleus and nucleon isospins. Particularly, for the $^5_\Lambda$He hypernucleus decay we have
\be
\tilde{{V}}_{N}=\binn{\tilde{{V}}_{0}\hspace{2.85cm}\mbox{for $N=n$}}{\tilde{{V}}_{0}+\frac{\tilde{V}_1}{2A}\hspace{2cm}\mbox{for $N=p$}},
\label{10}\ee
and for $^{12}_\Lambda$C decay we can write
\be
\tilde{{V}}_{N}=\binn{\tilde{{V}}_{0}-\frac{\tilde{V}_1}{2A}\hspace{2cm}\mbox{for $N=n$}}{\tilde{{V}}_{0}+\frac{\tilde{V}_1}{2A}\hspace{2cm}\mbox{for $N=p$}}.
\label{11}\ee
Thus, performing the replacements $\pb\rightarrow\alpha_N\pb$ and $\Pb\rightarrow\alpha_N\Pb$ in the matrix element given in \rf{45}, we can evaluate the effect of FSI on the asymmetry parameter.
Finally, in order to compare our theoretical results with experimental
data for the coincidence number ratio, we evaluate the number of emitted $nN$-pairs as
(remember the replacement $\Gamma$ by $\tilde{\Gamma}$ in Eq. \rf{e''}):
\be
N_{nN}=\frac{\tilde{\Gamma}_{N}}{\tilde{\Gamma}_{n}+\tilde{\Gamma}_{p}},
\label{17}\ee
being
\br
\tilde{\Gamma}_{N}&=&\int d\Omega_{p_n}\int d\Omega_{p_N}\int dF\sum_{s_ns_Nt_nt_NM_F}\label{18}\\
&\x&|((\alpha_N\pb_n)s_nt_n(\alpha_N\pb_N)s_Nt_N;J_FM_F|V|J_IM_I)|^2.\nn
\er
Based on this definition, it is important to remark here that our evaluation of the number of pairs will always satisfy the condition $N_{nn}+N_{np}=1$, which is only satisfied by the primary decay pairs in other theoretical  evaluations \cite{Bau07}. This is a reasonable result because our eikonal model only considers the scattering process of the final outgoing nucleons, without the possibility of producing additional secondary particles. In this way, the average number of detected pairs can not be different from those in the primary decay.

\section{Numerical results and discussion}

We have arrived to a very simple and easily manageable procedure which allows to evaluate
the observables $N_{nN}$ and $a_\Lambda$ including the minimum effect of the nucleon-nucleus interaction originated in
the scattering process of the emerging particles.
This method will be used here to evaluate the mentioned observables for $^5_\Lambda$He and $^{12}_\Lambda$C decays.

The values of the potentials $\tilde{V}_0$ and $\tilde{V}_1$ typically range from $10$ to $30$ MeV and $70$ to $110$ MeV, respectively \cite{Boh68}, and they are determined from analysis of a wide variety of data corresponding to neutron and proton
interactions with many different nuclei and at different incident energies. In Table \ref{tab1} we indicate the values assumed for these parameters and show the results obtained for NMHD observables for $^5_\Lambda$He and $^{12}_\Lambda$C decays.
\footnote{Note that the values used for these potentials are roughly within the range of validity for the eikonal approach, determined by the condition $\tilde{{ V}}_{N}/(<p>^2/2m_N)<<1$.}
We remark here that the values of $\tilde{V}_1$ are assigned from the fitting to the cross section of direct (pn) processes producing the isobaric analog state in the target nucleus \cite{Boh68}. This process correspond to a particle-particle + hole-hole scattering degree of freedom, as can be schematized by the first diagram in Fig. 2 a). Nevertheless, when an ejected proton  is converted in a neutron by the  interaction with the nucleons in the residual core, we can also have a process as shown in Fig. 2 b). To take into account this process, at least effectively,  with
the coordinate independent isovector operator we have assumed would be preferable to choose the maximum value for the strength $\tilde{V}_1/2A$. For completeness, we report in Table \ref{tab2} the experimental data.

\begin{table}
\begin{center}
\caption{Numerical results for the observables of $^5_\Lambda$He
and $^{12}_\Lambda$C hypernuclei.}
\label{tab1}
\renewcommand{\tabcolsep}{0.15pc}
\renewcommand{\arraystretch}{0.8}
\begin{tabular}{c|c|c|c|c|c|c}
\hline \hline
&$\tilde{V}_{0}$ (MeV)&$\frac{\tilde{V}_{1}}{2A}$ (MeV)&$N_{nn}$&$N_{np}$&$\frac{N_{nn}}{N_{np}}$&$a_\Lambda$\\
\hline\hline
$^5_\Lambda$He&&&&&&\\
\hline \hline
{\underline{no FSI}}&$ $&$ $&$ $&$ $&$ $&$ $\\
&$0$&$0$&$0.199$&$0.801$&$0.248$&$-0.538$\\
\hline
{\underline{with FSI}}&$ $&$ $&$ $&$ $&$ $&$ $\\
&$15$&$0$&$0.193$&$0.806$&$0.239$&$-0.510$\\
&$15$&$8.75$&$0.238$&$0.762$&$0.312$&$-0.489$\\
&$15$&$13.75$&$0.267$&$0.733$&$0.365$&$-0.477$\\
\hline \hline
$^{12}_\Lambda$C&&&&&&\\
\hline \hline
{\underline{no FSI}}&$ $&$ $&$ $&$ $&$ $&$ $\\
&$0$&$0$&$0.175$&$0.825$&$0.212$&$-0.530$\\
\hline
{\underline{with FSI}}&$ $&$ $&$ $&$ $&$ $&$ $\\
&$20$&$0$&$0.165$&$0.835$&$0.198$&$-0.500 $\\
&$20$&$3.18$&$0.192$&$0.808$&$0.238$&$-0.495$\\
&$20$&$5$&$0.210$&$0.790$&$0.265$&$-0.492$\\
\hline\hline \end{tabular} \end{center}
\end{table}

\begin{figure}[h!]
\vspace{-3.5cm}
\begin{center}
\includegraphics[height=12.cm,width=9.5cm]{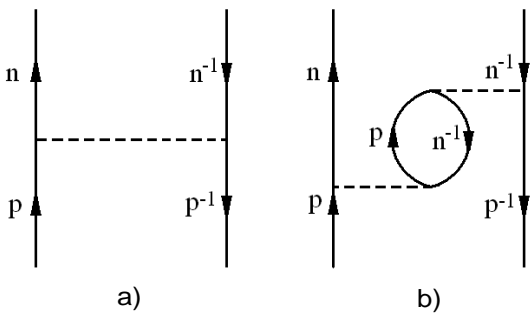}
\end{center}
\vspace{-4.cm}
\label{fig:1}
\caption{Graph a) schematizes particle-particle + hole-hole residual charge exchange interaction present in direct (pn) reactions producing the isobar analog state in the target nucleus. Graph b)
shows one of the mechanisms by which excited states in the residual nucleus can be generated.}
\end{figure}

\begin{table}
\begin{center}
\caption{Experimental data for NMHD observables.}
\label{tab2}
\renewcommand{\tabcolsep}{0.15pc}
\renewcommand{\arraystretch}{0.8}
\bigskip
\begin{tabular}{c|c|c}
\hline \hline
&$\frac{N_{nn}}{N_{np}}$&$a_\Lambda$\\
\hline
$^{5}_\Lambda$He&$-$&$0.07\pm 0.08^{+0.08}_{-0.00}$ \cite{Bhang05,Mar06}\\
&$0.45\pm 0.11\pm 0.03$ \cite{Outa05,Kang06}&$0.11\pm 0.08\pm 0.04$ \cite{Outa05,Mar05}\\
\hline
$^{12}_\Lambda$C&$-$&$-0.20\pm 0.26\pm 0.04$ \cite{Outa05,Mar05}\\
&$0.40\pm 0.10$ \cite{Bahn04,Kim06}&$- $\\
\hline\hline \end{tabular} \end{center}
\end{table}

Our results for the ratio $N_{nn}/N_{np}$ agree with theoretical and more elaborated estimations performed within microscopic models, which give values in the range $0.2-0.4$ \cite{Gar03,Gar04,Bau07}. Particularly, the more recent evaluation for $^{12}_\Lambda$C within the same $ \{ 2p1h \}$ final state space and within the ring approximation from Ref. \cite{Bau07} gives $N_{nn}=0.216$, $N_{np}=0.784$  $N_{nn}/N_{np}=0.275$ when no FSI are considered,
and $N_{nn}=0.253$, $N_{np}=0.906$ and $N_{nn}/N_{np}=0.279$ when their effect is included. Comparison with our results from Table 1 indicate a good agreement. Nevertheless the fact that in the mentioned reference $N_{nn} + N_{np} =0.253 + 0.906 = 1.159 >1 $, it is hard to understand.
This indicates that the theoretical frame described in the previous section to get $N_{nN}$ and the optical model eikonal approximation
to introduce the perturbative effects that mix different $2p1h$ configurations, provides a consistent  scheme to introduce the minimum effect of nucleon-nucleus interaction producing the scattering of the emerging particles.
From Tables 1 and 2 we also see that the effect of those interactions bring the theoretical value of the coincidence number ratio closer the experimental one. In fact, our results clearly indicate that $N_{nn}/N_{np}$ is increased mainly due to the effect of the isospin dependent part of the interaction, which allows the mixture between both channels. This leads to a difference between neutron and proton effective potentials proportional to $\tilde{V}_1/2A$ (see Eqs. \rf{10} and \rf{11}) which enforces the difference between both decay channels.

Additionally, the values obtained for the asymmetry parameter are in qualitative agreement with the ultimate theoretical evaluations \cite{Alb04}
in the sense that they exhibit the same tendency to reduce the magnitude of $a_\Lambda$, but can not revert the sign in the $^5_\Lambda$He case: for $^5_\Lambda$He ($^{12}_\Lambda$C) decay the authors obtain $-0.590$ ($-0.698$) when FSI effects are neglected and $-0.401$ ($-0.340$) when they are included. Indeed, FSI effects are not able to reproduce adequately the data for this observable, in spite that they increase the value. The changes are mainly caused by the combined effect of $\tilde{V}_0$ and $\tilde{V}_1$ parts of the optical potential, which modify the PC and PV matrix elements in a different way, through the momenta replacement $\pb\rightarrow \alpha_p \pb$ and $\Pb\rightarrow \alpha_p \Pb$ (see Eq. \rf{7}). We note that, as $a_\Lambda$  is only measured for protons and it is theoretically determined by a ratio between quantities calculated both for protons (see Eq. \rf{a}) the charge exchange effects of FSI are not so important in this case. Thus, one must infer that new modifications to the model, as for example the inclusion of two-pion correlated or $a_1$-meson exchange in the $N\Lambda$ potential \cite{Chu0708,Ito08}, could be still carried out to achieve a better description of NMHD.

As a final remark, if $2p2h$ GSC were included, the primary decay rate $\Gamma_N$  should be affected by a factor $|c_0|^2$ (see Eq. \rf{efg}) while the pair numbers $N_{nN}$ calculated within the $2p1h$ subspace remain unaffected. The 2N mode should be taken into account with amplitudes $c_0 c_2$
and the corresponding contributions of this mode should be added to each observable $\Gamma_T$, $N_{nN}$, $\cdots$, etc. However, caution must be taken in the way of evaluating $c_0$ and $c_2$. In fact, if a first order perturbative approximation is taken for them, the prescription \rf{efgh'} must be assumed. Otherwise,
if one wish to introduce second order normalization effects, all second order effects must be included in the ground state, which implies to introduce $4p4h$ configurations.

\section{Summarizing conclusions}

We have analyzed expressions for the evaluations of the number of pairs $N_{nN}$ and the asymmetry parameter $a_\Lambda$ when FSI are considered in leading order. Working at a first step within the $2p1h$ configuration subspace, we have included in a simple way the effect of the scattering of the outgoing nucleons due to the residual core through the optical model eikonal approximation within a finite nuclei framework.
Our approach uses only two parameters: the isoscalar and isovectorial strengths of the optical potential.
Present evaluation does not pretend to replace more elaborated formalisms based on numerical codes or microscopical considerations, mainly because they contribute to the physical insight of the process, but to provide a simple and nice alternative as an application of the well known eikonal method. We have arrived to a very simple analytical procedure (see Eq. \rf{7}) which can be implemented independently of the model adopted for the description of NMHD.
Our results show that the effect of the isovector interaction on the ratio $N_{nn}/N_{np}$ is stronger than on the asymmetry $a_\Lambda$ because: i) in the first case the modification is originated from a difference between neutron and proton potentials due mainly to the isovector part of the optical nuclear potential allowing the possibility of charge exchange; ii) in the second one, being $a_\Lambda$ measured only for proton induced decay, modifications due to final interactions are incorporated only by means of the differences produced between PC and PV matrix elements through the momenta renormalization \rf{7}.

\bigskip

{\bf Acknowledgments}

\bigskip

The authors are grateful to Eduardo Bauer for fruitful discussions. C.B. and A.M. are fellows of the CONICET, CCT La Plata (Argentina) and thank the partial support under grant PIP 06-6159. They also thank to the Agencia Nacional de Promoci\'on Cient\'{\i}fica y Tecnol\'ogica for supporting in part this work through PICT-2007-00861. C.B. thanks to the Brazilian Ministerio da Ciencia e Tecnolog\'{\i}a for partial supporting of this research. S.B.D. thanks to Brazilian agency CNPq for partial support.

\end{document}